\documentclass[%
 reprint,
 amsmath,amssymb,
 aps,
prd
]{revtex4-2}

\usepackage[colorlinks=true
,urlcolor=blue
,anchorcolor=blue
,citecolor=blue
,filecolor=blue
,linkcolor=blue
,menucolor=blue
,linktocpage=true
,pdfproducer=medialab
,pdfa=true
]{hyperref}

\usepackage{graphicx}% Include figure files
\usepackage{dcolumn}% Align table columns on decimal point
\usepackage{bm}% bold math
\usepackage{ulem}
\newcommand\aap{A\&A}                % Astronomy and Astrophysics
\newcommand\apjl{ApJ}                % Astrophysical Journal, Letters
\newcommand\jcap{J.~Cosmology Astropart. Phys.} % Journal of Cosmology and Astroparticle Physics
\newcommand\mnras{MNRAS} 
\newcommand\sovast{Soviet Astronomy}           
\newcommand\physrep{Physics Reports}

\begin{document}

\title{Free-free background radiation from accreting primordial black holes}
\author{Hiroyuki Tashiro}
\email{hiroyuki.tashiro@nagoya-u.jp}
\author{Katsuya T. Abe}
\author{Teppei Minoda}
\affiliation{Division of Particle and Astrophysical Science,
Graduate School of Science, Nagoya University,
Chikusa, Nagoya 464-8602, Japan}

\date{\today}% It is always \today, today,

\begin{abstract}
Baryonic gas falling onto a primordial black hole~(PBH) emits photons via the free-free process. 
These photons can contribute the diffuse free-free background radiation in the frequency range of the cosmic microwave background radiation~(CMB).
We show that the intensity of the free-free background radiation from PBHs depends on the mass and abundance of PBHs.
In particular, considering the growth of a
dark matter~(DM) halo around a PBH by non-PBH DM particles strongly
enhances the free-free background radiation.
Large PBH fraction increase the signal of the free-free emission.
However, large PBH fraction also can heat the IGM gas and, accordingly, 
suppresses the accretion rate.
As a result, the free-free emission decreases when the PBH fraction is larger than 0.1.
We find that the free-free emission from PBHs in the CMB and radio frequency is much lower than the CMB blackbody spectrum and
the observed free-free emission component in the background radiation.
Therefore, it is difficult to obtain the constraint from the free-free emission observation.
However further theoretical understanding and observation on the free-free emission from cosmological origin 
is helpful to study the PBH abundance with the stellar mass.
\end{abstract}

\maketitle
\section{Introduction}
%The standard $\Lambda$CDM cosmology has achieved remarkable success so far.    
%Many different cosmological observations have been conducted until now
%and their observation results are consistent 
%with the prediction from the standard $\Lambda$CDM cosmology.
%However this success reveals that the
%evolution of the structure in the Universe
%is strongly affected by unknown matter, called dark matter~(DM).

The existence of dark matter~(DM) is strongly supported by a variety of astrophysical and cosmological observations~\cite{2006ApJ...648L.109C, 2020A&A...641A...6P}.
The most explored candidate is new elementary particles beyond the standard model of particle physics, i.e.,
weakly interacted massive particle~\cite{1996PhR...267..195J,2005PhR...405..279B}.
Although many experiments have been conducted aggressively
to study these candidates,
the detection has not been reported yet.

A primordial black hole~(PBH)
is one of the viable candidates
for non-particle DM.
The original idea of PBH is advocated by
Zel'dovich and Novikov in 1960~\cite{1967SvA....10..602Z}.
PBHs form from the gravitational collapse
of the overdense regions in the early universe,
and their mass is in a wide range from the Planck mass
to masses much larger than the solar mass~\cite{1971MNRAS.152...75H}.
PBHs are considered to be a candidate of DM
for a long time~\cite{1974MNRAS.168..399C,1975Natur.253..251C}. 
%Since many people confirmed the particle-type dark matter at that time, PBHs were lesser-studied as the nature of DM. However, 
Although, currently, there is no evidence for the existence of PBHs, 
the recent detections of gravitational waves by LIGO/VIRGO
draw attention to PBHs~\cite{2016PhRvL.116t1301B}.
From the existing constraints,
PBHs with mass responsible for the above GW events
with stellar-mass BH binary merger,
cannot account for the total amount of DM.
However, still, PBHs can be a subdominant component of DM
\cite{2020ARNPS..70..355C,2021JPhG...48d3001G}.

The stringent constraints on PBHs
with mass $M\gtrsim \mathcal{O}(1)M_{\odot}$
have been obtained in the gas accretion scenario onto PBHs
\cite{2008ApJ...680..829R,2017PhRvD..95d3534A}. 
PBH gravitational force attracts the surrounding gas.
When accreting onto a PBH,
gas is heated up and ionized enough to
emit X-ray and UV photons by the free-free emission.
These high energy photons can modify
the thermal and ionization history of the diffuse background gas 
before the epoch of reionization.
Since such modification affects the CMB anisotropies,
CMB observations can provide the strong constraint
on the amount of PBHs
\cite{2017PhRvD..95d3534A,2017PhRvD..96h3524P,2020PhRvR...2b3204S}.
The measurement of these effects has been also discussed
in the Sunyaev-Zel'dovich effect fluctuations~\cite{2019PhRvD..99j3519A} and future 21-cm observations\cite{2013MNRAS.435.3001T,2017JCAP...08..017G,2018PhRvD..98b3503H,2019PhRvD.100d3540M}.

The free-free emission from accreting gas
has a continuous spectrum below the UV frequency.
Since these photons are not consumed for
the heating and ionization of the background gas,
we can observe this
% these
emission directly as diffuse background radiation.
According to detailed studies in the CMB frequency range,
the radiation with the free-free spectrum is
in the order of several tens of $\mu$K
and most of them are considered to be of Galactic origin
\cite{2011ApJ...734....6S,2016A&A...594A..10P}.
However, some fraction of them could be the contribution 
from large-scale structure or the structure formation in the early universe
\cite{2004ApJ...606L...5C,2011MNRAS.410.2353P,2019MNRAS.486.3617L,abeetal2021}.

In this paper, we evaluate the intensity of the free-free background radiation from PBHs.
We also compare it with the current observational data of the free-free emission from the CMB observation.
Furthermore, we discuss the impact of the free-free emission on the 
radio low frequency range.
Throughout this paper,
we take the flat $\Lambda$CDM model with the cosmological parameters
$(\Omega_{\rm m},\Omega_{\rm b},h)$
=$(0.3,0.05,0.7)$.

\section{free-free emission around a PBH}
\label{sec:acc}

A PBH can accrete baryon gas after the matter-radiation equality. During the accretion, the gas is heated up and
ionized. As a result, the accreting gas around a PBH can emit radiation via a free-free emission process.
In order to evaluate this  free-free emission, it is required to obtain the gas profile of the density, temperature, and ionization rate around a PBH. 
Recently Refs.~\cite{2008ApJ...680..829R,2017PhRvD..95d3534A}
have investigated the gas accretion onto a PBH in the cosmological context.
Assuming the spherical accretion and the steady-state approximation,
they have obtained a simple model of gas profile. 
In this section, following the procedure in Ref.~\cite{2017PhRvD..95d3534A},
we evaluate the free-free emission from a PBH.

We consider the accretion gas onto a PBH
with mass $M_{\rm PBH}$ at a redshift $z$.
The simple spherical accretion can be described
by the Bondi accretion~\cite{1952MNRAS.112..195B}.
Based on this model,
we introduce the parameter~$\lambda$ describing the spherical accretion rate $\dot{M}$,
\begin{align}
\dot{M} =4 \pi \lambda \overline \rho_{\rm b} r_{\rm B}^2 c_{\rm s},
\end{align}
where $\overline {\rho}_b $ is the background baryon gas density,
$c_{\rm s}$ is the sound velocity of the background baryon gas, and $r_{\rm B}$ is the Bondi radius, $r_{\rm B}= GM_{\rm{PBH}}/c_{\rm s}^2$.

In the determination of the radial profiles of the accreting gas,
one of the
largest theoretical uncertainties is the ionization process of the gas.
The gas ionization is accomplished through not only the collisional ionization
% and
but also photoionization by the emission
from hot ionized gas in the central region.
We assume that the gas is ionized only by the thermal collision ionization process. In this assumption, for simplicity, the ionization does not proceed until the temperature reaches the critical temperature $T_{\rm ion } \sim 10^4~\rm K$.
% Then ionization fraction increases with keeping $T=T_{\rm ion}$ until $x_e =1$.
When reaching the critical temperature, the ionization fraction would increase until $x_e =1$ with keeping $T=T_{\rm ion}$.
This assumption provides the conservative estimation of the free-free emission~\cite{2017PhRvD..95d3534A}.

% If the emission is enough strong,
% it enhances the ionization process and heats the gas.
% On the other hand, when the emission is weak,
% the ionization is accomplished by the collisional ionization.
% Since the thermal energy is consumed in the ionization energy,
% the gas temperature does not increase during the ionization process.

% The conservative estimation of the free-free emission is obtained by considering only thermal collision.
% In this case, the ionization does not proceed until the temperature reach the critical temperature $T_{\rm ion } \sim 10^4~\rm K$.
% Then ionization proceeds with keeping $T=T_{\rm ion}$ until $x_e =1$.

Ref.~\cite{2017PhRvD..95d3534A} divides 
the radial gas profile into three regions; the outermost region, collisional ionization region, and innermost adiabatic region.
In the outermost region, the gas density and the temperature increases from the background values, $\overline{\rho }_{\rm b}$ and $\overline{T}$, as the radius, $r$, decreases.
When the steady-state approximation is valid
%(but the accretion time scale is shorter than the cosmological one),
the spherical accreting gas can be described by the following equations,
\begin{align}
&4 \pi r^2 \rho_{\rm b} |v| = \dot{M},
\label{eq:mass_con}
\\
&v \frac{d v}{d r} = - \frac{G M_{\rm PBH}}{r^2}  - \frac1{\rho_{\rm b}} \frac{d P}{dr} - \frac{4}{3} \frac{\overline{x}_e \sigma_{\rm T} \rho_{\rm CMB}}{m_{\rm p} c} v, 
\label{eq:euler}
\\
&v \rho_{\rm b}^{2/3} \frac{d}{dr}\left( \frac{T}{\rho_{\rm b}^{2/3}}\right) = \frac{8 \overline{x}_e \sigma_{\rm T}\rho_{\rm CMB}}{3 m_e c (1 + \overline{x}_e)}(T_{\rm CMB} - T), 
\label{eq:heat-eq}
\end{align}
where $\rho_{\rm b}$ is the gas energy density, $|v|$ is the gas radial velocity, $P$ is the gas pressure, $\rho_{\rm CMB}$ is the CMB energy density, $\sigma_{\rm T}$ is the cross-section of the Thomson scattering, and $\overline{x}_{\rm e}$ is the background ionization rate.
The first equation is the mass conservation equation, and the second and the third equations are the momentum equation and heat equation of gas, respectively.
The last terms in Eqs~\eqref{eq:euler} and~\eqref{eq:heat-eq}
represent the Compton drag and cooling terms. They provide a significant impact on the accretion in high redshifts.
It is useful to introduce two dimensionless parameters describing the time scales of these CMB effects,
\begin{align}
\beta = \frac{4}{3} \frac{\overline{x}_e \sigma_{\rm T} \rho_{\rm CMB}}{m_{\rm p} c} t_{\rm B},\quad
\gamma = \frac{8 \overline{x}_e \sigma_{\rm T}\rho_{\rm CMB}}{3 m_e c (1 + \overline{x}_e)} t_{\rm B},
\end{align}
where $t_{\rm B}$ is the time scale of the Bondi accretion,
$t_{\rm B} = GM_{\rm PBH}/c_{\rm s}^3$.

Once $\lambda$ is determined, the gas profiles in the outermost region
are provided by Eqs.~\eqref{eq:mass_con}-\eqref{eq:heat-eq}.
Although the accretion rate depends on the physical condition around a PBH, Ref.~\cite{2017PhRvD..95d3534A} have found out $\lambda$
which can provide the physically valid solution of Eqs.~\eqref{eq:mass_con}-\eqref{eq:heat-eq}.
The approximated form of such $\lambda$ is given by
\begin{align}
\lambda(\gamma, \beta)  =
%\frac{\lambda(\gamma; \beta \ll 1) \lambda(\gamma \gg 1; \beta)}{\lambda_{\rm iso}}. 
%\left(\lambda_{\rm ad} + (\lambda_{\rm iso}-\lambda_{\rm ad})\left(\frac{\gamma^2}{88+\gamma^2} \right)^{0.22}\right)
\frac{\lambda_{\rm ad} + (\lambda_{\rm iso}-\lambda_{\rm ad})\left(\frac{\gamma^2}{88+\gamma^2} \right)^{0.22}
}{(\sqrt{1+\beta} +1)^2}
\exp\left[\frac{9/2}{3+\beta^{3/4}}\right] ,
\label{eq:lambda-full}
\end{align}
where $\lambda_{\rm ad } = (3/5)^{3/2}/4 
%\approx 0.12 
$ and 
$\lambda_{\rm iso } = {\rm e}^{3/2}/4 
%\approx 1.12 
$.
In this case, following Eqs.~\eqref{eq:mass_con}-\eqref{eq:heat-eq},
the gas density and the temperature are proportional to
$T\propto \rho^{2/3}\propto 1/r$, respectively.
The smallest
radius of the outermost region is $r_{\rm i}$ where the temperature reaches $T=T_{\rm ion}$.
The radius $r_{\rm i}$~is $ r_{\rm i} \approx f T_{\rm{CMB}}r_{\rm{B}}/T_{\rm{ion}} $, where $f$ the coefficient of $0.3$ which can be smaller when Compton cooling is effective.
%In this region, the collisional ionization is not effective. The ionization fraction is the same as the background value, $x_{e} = \overline{x}_e$. \KA{??: After the decoupling, $\overline{x}_e$ rapidly drop down. Therefore, baryon gas in this region does not contribute the free-free emission.}

Inside $r<r_{\rm i}$, the collisional ionization becomes effective.
In this collisional ionization region, as we mentioned above, the gas keeps the temperature, $T_{\rm ion}$, with increasing the ionization fraction.
% This is because the cooling due to the increment of free particle number and the energy loss by the proceeding of the ionization is compensated by the adiabatic compression heating of the gas.
% This compensation provides the relation between the gas density and ionization fraction.
However, since the gas suffers the adiabatic compression,
the gas density follows $\rho \propto r^{-3/2}$.
Then, the ionization fraction is given in $(1+x_e) \propto r^{-1/8}$~\cite{2017PhRvD..95d3534A}.
The radius where $x_e = 1$ is provided in $r_{\rm e} = (1+\overline{x}_e)^8 r_{\rm i} /2^8$.

In the innermost adiabatic region, $r < r_{\rm e}$, the gas density and temperature follow the adiabatic compression. The gas density and the temperature are proportional to $\propto r^{-2/3}$ and $r^{-1}$, respectively. When the temperature exceeds the electron mass, relativistic correction is required. As a result, the temperature profile is modified to $T \propto r^{-2/3}$ in the region where $T > m_ec^2$.
The accretion flow reaches the Schwarzschild radius, $r_{\rm S}$,
and the gas velocity becomes almost the speed of light.
At $r_{\rm S}$, the temperature of gas reaches $T \sim 10^{9} \rm K$.

%At $r_{\rm S}$, the gas density is estimated in
%\eq{
%\rho_{\mathrm{S}} %&=\frac{\lambda}{\left(c / c_{\m{s}}\right)\left(r_{\mathrm{S}} / c_{\m{s}}\right)^{2}} \rho_{\infty}=\frac{\lambda}{4\left(c_{\m{s}} / c\right)^{3}} \rho_{\infty} \\ 
%=\frac{\lambda}{4}\left(\frac{m_{p} c^{2}}{\left(1+\overline{x}_{e}\right) \overline{T}}\right)^{3 / 2} \overline{ \rho}_{\rm b}.
%}
% \begin{align}
% \rho_{\rm S}
% \end{align}

So far we consider the spherical Bondi accretion.
However, there exists the relative velocity between
DM and baryons.
The coherent scale of the velocity is several comoving Mpc.
The amplitude is larger than the sound speed of baryon gas soon after the recombination epoch, $\langle v_L^2 \rangle \approx {\rm min}[1, z/10^3] \times 30 ~{\rm km/s}$ as shown in Ref.~\cite{2010PhRvD..82h3520T}.
Due to this relative velocity,
the spherically accretion model must be corrected.
Ref.~\cite{2017PhRvD..95d3534A} proposed the replacement of $c_{\rm s}$ in the above discussion to 
\begin{align}
v_{\rm eff} 
\approx \begin{cases} \sqrt{v_{\rm B} \langle v_{\rm L}^2 \rangle^{1/2}}, \ \ &v_{\rm B} \ll \langle v_{\rm L}^2 \rangle^{1/2},\\
v_{\rm B}, \ \ &v_{\rm B} \gg \langle v_{\rm L}^2 \rangle^{1/2}.
\end{cases} \label{eq:veff}
\end{align}
Following this procedure, we calculate the radial profile in the subsequent sections.

\begin{figure}
  \centerline{
  \includegraphics[scale=0.6]{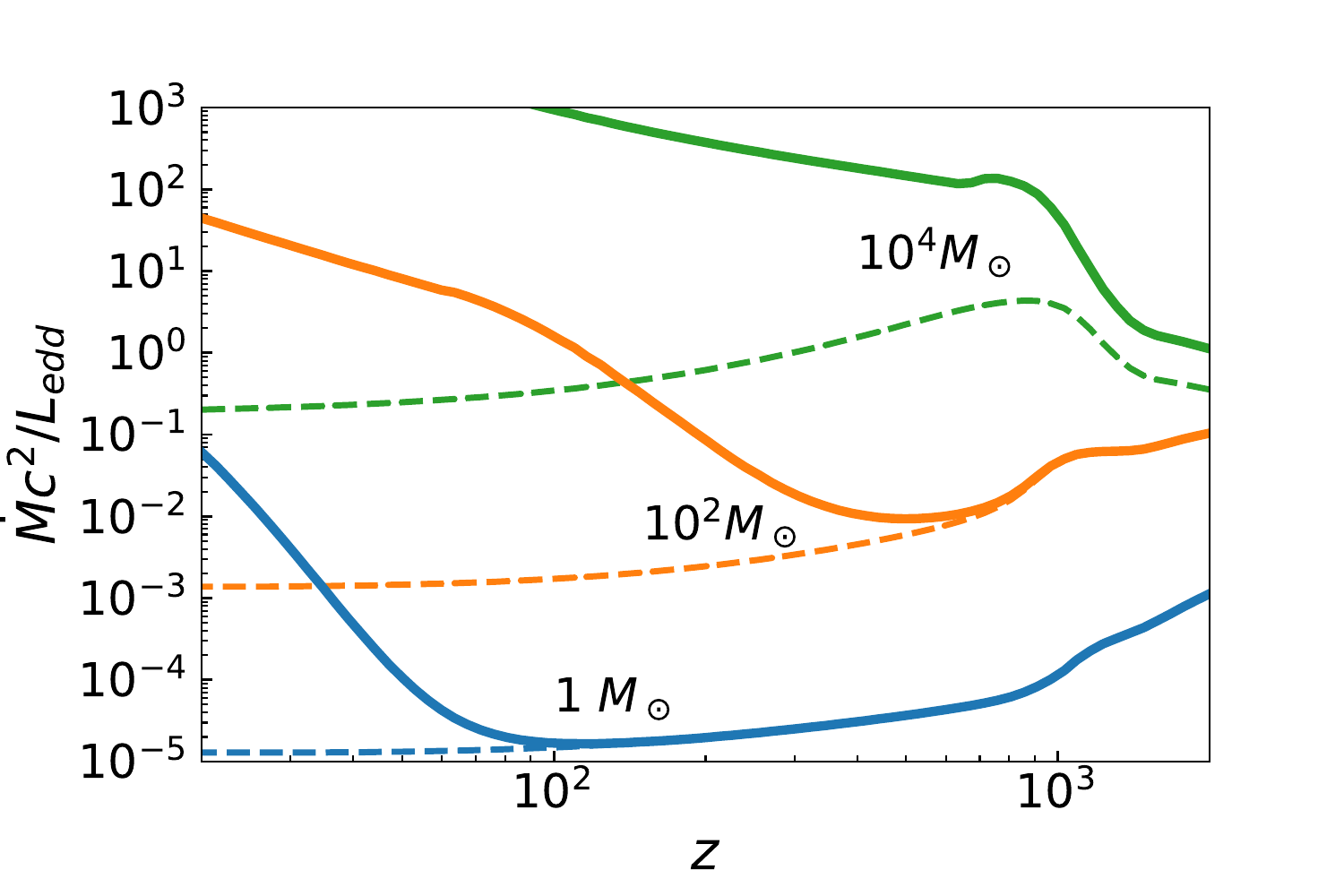}
  }
  \caption{Accretion rate normalized in the Eddington luminosity. The solid lines are with the DM halo effect while the dashed lines are without the DM halo effect. 
  From top to bottom, the lines represents for $M_{\rm PBH} =10^4$,~$10^2$ and $1~M_\odot$.}  \label{fig:mdot}
\end{figure}

Although we consider the accretion of baryon gas onto PBHs,
due to the gravity of a PBH, non-PBH DM particles also accrete on the PBH, and a DM halo forms around the PBH. The created PBH halo enhances the accretion of baryons onto PBHs and, resultantly, amplifies the free-free emission.
In order to include this effect, we adopt the simple semi-analytical model provided in Ref.~\cite{2020PhRvR...2b3204S}.
In this model, the DM halo mass~$M_{\rm halo}$ grows as $M_{\rm halo } \propto (1+z)$ after the matter-radiation equality suggested in Ref.~\cite{2007ApJ...665.1277M}.
Fig.~\ref{fig:mdot} shows the enhancement of the accretion rate by considering the DM halo effect.
In this plot, we normalize the mass accretion to the Eddington luminosity.
The solid lines represent the ones with the DM halo effect while the dashed lines are without the DM halo effect.
The DM halo enhancement is prominent in lower redshifts.
For example, in $z\approx 30$, the total mass gravitationally attracting the baryon gas in the outermost region becomes 
100 times larger than the initial mass of the central PBH.
In lower redshifts, the gravity of growing DM halos dominates the PBH and strongly enhances the accretion of baryon gas.
The gas falling into a PBH increases the mass of the PBH.
We calculate the increment of the PBH mass after the matter-radiation equality, $\Delta M(z) = \int _{t_{\rm eq}} ^{t(z)}  \dot M ~dt$. 
%Fig.~\ref{fig:deltaM} shows the evolution of the accretion mass in the term of $\Delta M/M_{\rm PBH}$. 
Because of the accretion, 
the PBH mass rapidly increases in low redshifts.
As the initial PBH is massive, the accreted mass grows PBH mass more.
The growth of the PBH mass increases the Schwartzshild radius more than the initial one.
However, this increase does not modify our final result for the PBH mass less than $10^4 M_\odot$.

%\begin{figure}
%  \centerline{
%  \includegraphics[scale=0.6]{PBH_deltaM.pdf}
%  }
%  \caption{Accreted mass after the matter-radiation equality. The solid lines are with the DM halo effect while the dashed lines are without the DM halo effect. 
%  From top to bottom, the lines represents for $M_{\rm PBH} =10^4$,~$10^2$ and $1~M_\odot$.}
%  \label{fig:deltaM}
%\end{figure}

\section{diffuse free-free background radiation}

Ionized hot accreting gas can emit radiation via free-free emission. First, we evaluate the intensity of the free-free emission from an individual PBH.

Now we consider a line of sight with impact parameter $b$ in unit of $r$ from the center of the PBH.
The intensity of the free-free emission at a frequency~$\nu$
is calculated by the integration over
this line of sight
\begin{align}
I_\nu(b) = \int_{-R_{\rm max}} ^{R_{\rm max}} f_\nu(R) j_\nu(\ell ) dR,
\end{align}
where $R$ is the position along the line of sight, and $R=0$ represents the center of this line of sight in the accreting gas.
The radial distance~$\ell$ satisfies $\ell^2 = R^2 + (b r_{\rm s})^2$, and $R_{\rm max}$ is given by $R_{\rm max}
= \sqrt{1-b^2} r_{\rm s}$.
The function $f_\nu (R)$ is given by
\begin{align}
f_{\nu}(R) = \exp[- (\tau_\nu (R_{\rm max} )
-\tau_\nu (R)   )],
\end{align}
where 
the optical depth $\tau_{\nu}(R)$ at the frequency $\nu$ is obtained from 
\begin{align}
\tau_{\nu}(R) = \int_{-R_{\rm max}} ^{R}  \alpha_\nu (\ell) d R,
\end{align}
with the absorption coefficient~$\alpha_\nu$.

The radial profiles of the emission and absorption coefficients
are given by~\cite{Radiative_process_in_Astrophysics}
\begin{align}
j_\nu(r) &= \frac{2^3  e^6}{3 m_e c^3} \sqrt{\frac{2 \pi}{3 k_{\rm B} T m_e}} n_e n_p \bar{g}_{\rm ff},
\\
\alpha_\nu(r) &= \frac{4 e^6}{3 m_e c k_{\rm B} T} \sqrt{\frac{2 \pi}{3 k_{\rm B} T m_e}} n_e n_p \bar{g}_{\rm ff},
\end{align}
where $e$ is the electric charge, and $\bar{g}_{\rm ff}$ is the velocity averaged Gaunt factor. We adopt the approximation form in Ref.~\cite{2011piim.book.....D},
\begin{align}
\bar{g}_{\rm ff} =
\log \left\{\exp \left[5.960-\sqrt{3} / \pi \log \left(\nu_{9}T_{4}^{-3 / 2}\right)\right]+\mathrm{e}\right\},
\end{align}
where $\nu_{9}\equiv \nu/(1~\rm{GHz})$, $T_{4} \equiv T/(10^4~{\rm K})$~and $\rm e$ is the Napier's constant.
When the frequency is much smaller than $k_B T /h$ and the absorption is negligible, 
the frequency dependence comes from only $\bar g_{\rm ff}$, and its dependency on the frequency is very weak.
As a result, the free-free intensity is almost scale-invariant, $I \propto \nu^{-0.1}$.

The intensity averaged over the cross-section is 
\begin{align}
\overline{I}(\nu) = \int_0 ^1 I_{\nu} (b)  b d b.
\end{align}
Since the radial profile of gas around a PBH depends on the redshift $z$,
the averaged intensity is also a function of $z$. Hereafter we represent it as $\overline{I}(\nu, z)$.

The intensity averaged over the full sky is obtained by taking into account PBHs with the comoving number density $n_{\rm PBH}$,
\begin{align}
I_{\rm PBH}(\nu_{\rm obs}) =  \frac{1}{4 \pi}
\int^\infty_{z_{\rm c}}
\frac{\overline{I}(\nu_z,z)}{(1+z)^3} \Delta \Omega(z) n_{\rm PBH} \frac{dV_{\rm c}} {dz} dz,
\label{eq:intensity_fullsky}
\end{align}
where $\nu_z = (1+z) \nu_{\rm obs}$, $\Delta \Omega(z)$ is the solid angle of a PBH at $z$, $\Delta \Omega = \pi (1+z)^2 r^2_s/D^2(z)$ with the comoving distance $D(z)$, and $V_c$ is the comoving volume of the Universe at the redshift $z$.
In the equation, $z_{\rm c}$ is the 
lower limit of the redshift integration.
Although we assume the adiabatic evolution of the background gas temperature, stars and galaxies can heat up it in lower redshift. The heated gas temperature suppresses the accretion rate.
The impact of the star and galaxy formation on the gas temperature evolution is still unknown.
In order to avoid 
this model uncertainty, 
we set $z_{\rm c}=20$ in this paper. 
% \KA{??: and we discuss the dependence of the intensity on $z_{\rm c}$ in the next section.}
Eq.~\eqref{eq:intensity_fullsky} tells us that the free-free emission intensity is proportional to $f_{\rm PBH}$, because the comoving number density $n_{\rm PBH}$ is 
$n_{\rm PBH} = f_{\rm PBH} \Omega_{\rm DM} \rho_{\rm c}/M_{\rm PBH}$.
When the optically thin approximation is valid~(the absorption contribution can be negligible),
Eq.~\eqref{eq:intensity_fullsky} is approximated to
\begin{align}
I_{\rm PBH}(\nu_{\rm obs}) \approx  \int  \frac{\overline{j}(\nu_z , z)}{(1+z)^4}  
\frac{dD(z)}{dz} dz,
\label{eq:intensity_thin}
\end{align}
where $\overline{j}$ is given by
\begin{align}
\overline{j}(\nu, z) = (1+z)^3 n_{\rm PBH}   \int_0^{r_s} j_\nu(r)  
4 \pi r^2 dr. 
\label{eq:j_thin}
\end{align}

\section{Constraint on the PBH abundance}

The free-free emission has an almost-flat frequency spectrum
as mentioned in the previous section. 
In the CMB frequency range, such frequency-invariant free-free emission is well studied as one of the main foreground sources~\cite{2016A&A...594A..10P}.
In order to compare the observed free-free signal,
we calculate the free-free emission induced by PBHs at the CMB frequency range.

The free-free emission can  produce high energy photons 
whose energy is high enough to
the heat and ionize the background IGM~\cite{2008ApJ...680..829R,2017PhRvD..95d3534A}.
Since the accretion rate depends on the ionization rate and temperature of the background IGM as mentioned in Sec.~\ref{sec:acc}.
In order to the ionization and heating due to the PBH emission, we follow the procedure in Ref~\cite{2017PhRvD..95d3534A}.
Modifying the public 
recombination code {\tt HYREC}~\cite{2010PhRvD..82f3521A,2011PhRvD..83d3513A},
we 
calculate the thermal evolution of the IGM with the free-free emission effect from PBHs
and, then,
evaluate the free-free intensity in the CMB frequency range.

\begin{figure}
  \centerline{
  \includegraphics[scale=0.6]{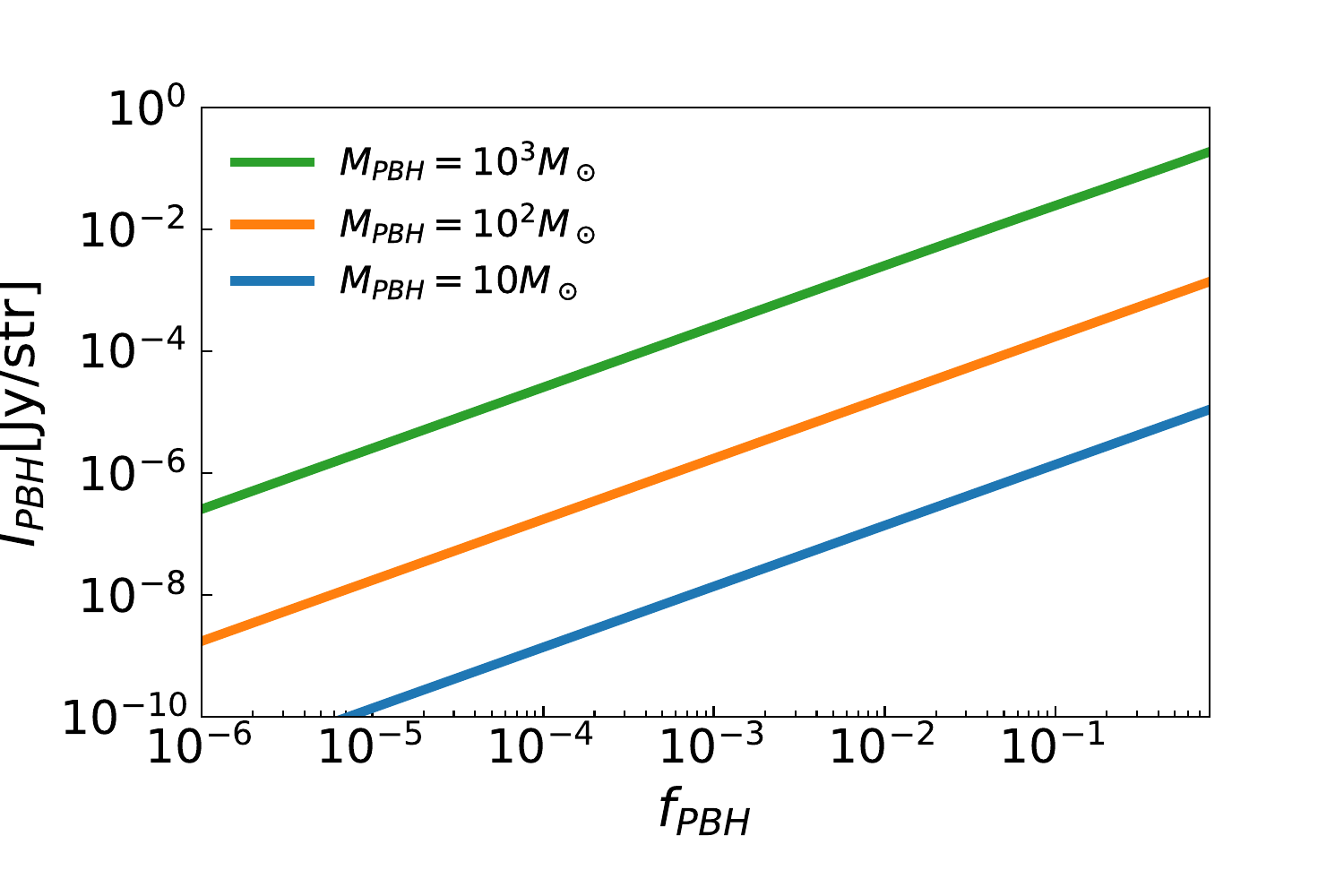}
  }
  \caption{Intensity of the diffuse free-free background radiation from PBHs. Here the DM halo effect is not included. The solid lines represent the intensity for $M_{\rm PBH} = 10^3~M_\odot$, $10^2~M_\odot$ and $10~M_\odot$ from top to bottoms.}
    \label{fig:intensity}
\end{figure}

Fig.~\ref{fig:intensity} shows the emission intensity at~the observation frequency, $\nu_{\rm{obs}}=44~$GHz.
% \KA{needed?: The intensity is proportional to $f_{\rm PBH}$.}
In this figure, we neglect the DM halo effect.
In this case, the modification
on the thermal history due to high energy photons from  PBHs does not affect the accretion rate. As the result, the intensity is simply proportional to $f_{\rm PBH}$ through the PBH number density in Eq.~\eqref{eq:intensity_fullsky}.

\begin{figure}
  \centerline{
  \includegraphics[scale=0.6]{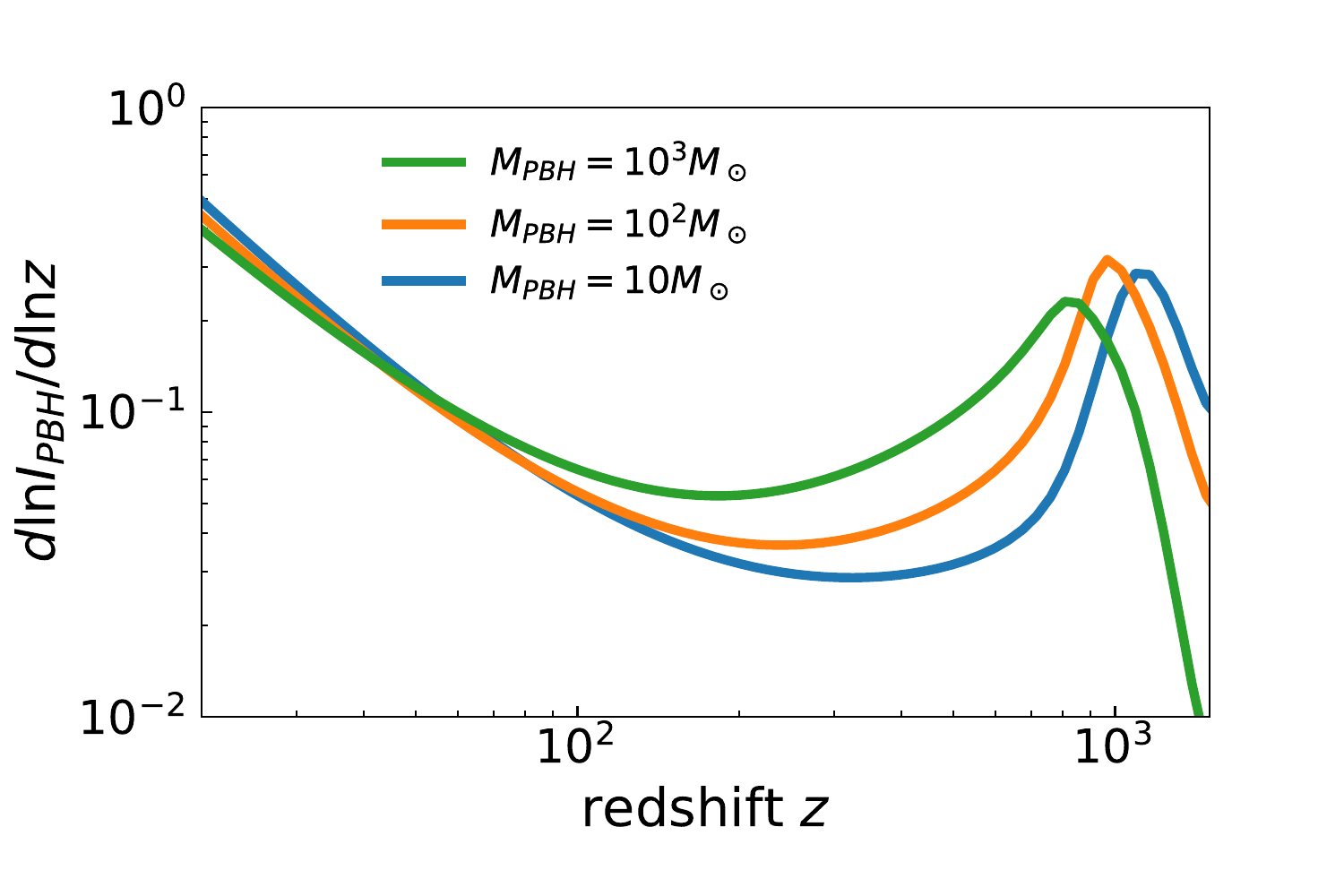}
  }
  \caption{Intensity of the diffuse free-free background radiation from PBHs. Here the DM halo effect is inclu. The solid lines represent the intensity for $M_{\rm PBH} = 10^3~M_\odot$, $10^2~M_\odot$ and $10~M_\odot$ from top to bottoms.}
    \label{fig:redshift}
\end{figure}

We plot the redshift distribution of the free-free intensity in Fig.~\ref{fig:redshift}.
The free-free emission from high redshifts is suppressed due to the cosmological redshifts.
Therefore, the contribution from low redshift becomes large.
One can find that 
the non-negligible contribution arises around the recombination epoch 
This is because the intensity depends on the accretion rate. The accretion rate is high as the redshift increases and has the peak around the recombination epoch.
Therefore, the intensity distribution also has a peak shape around the recombination cannot negligible.

\begin{figure}
  \centerline{
  \includegraphics[scale=0.6]{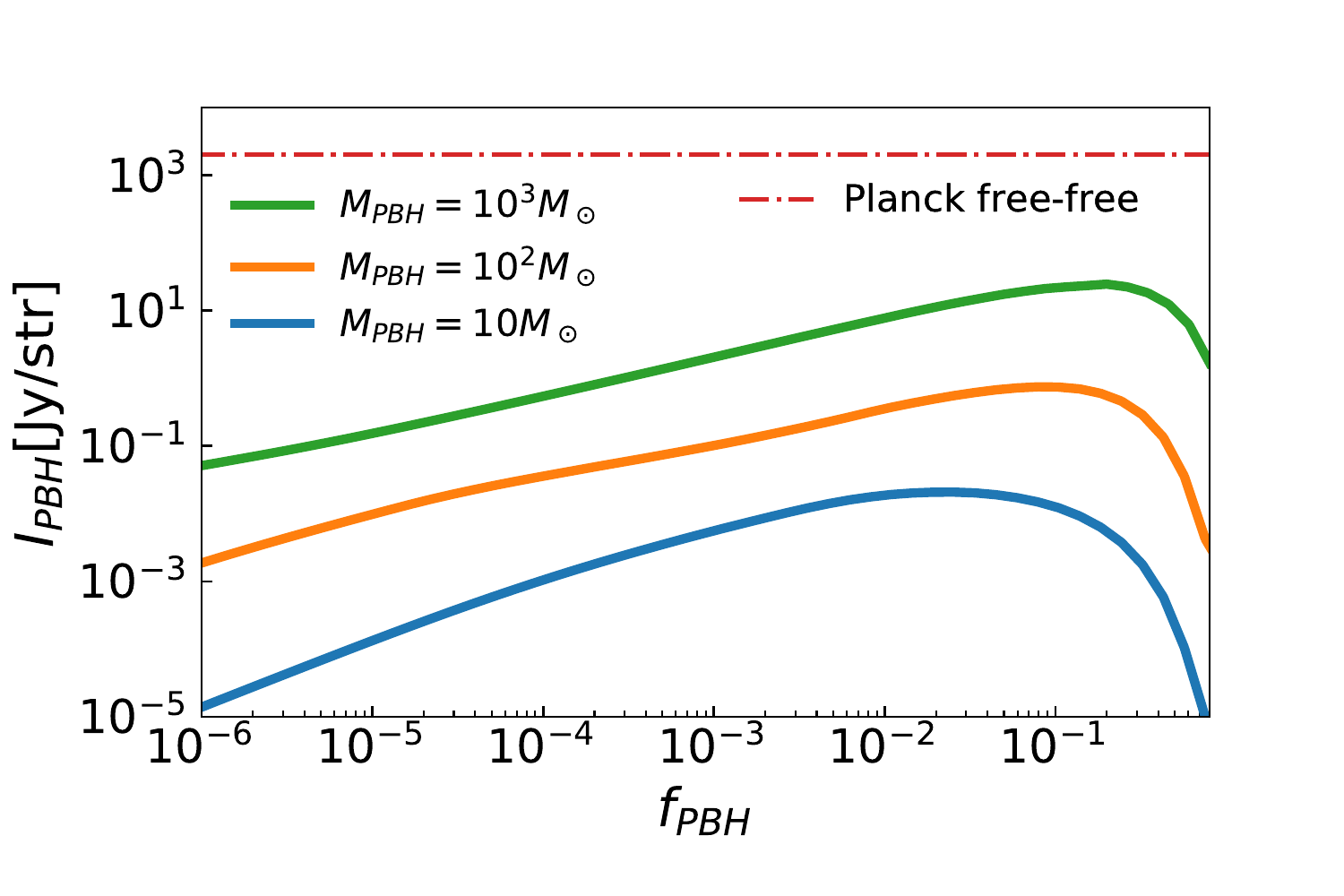}
  }
  \caption{Intensity of the diffuse free-free background radiation from PBHs. Here the DM halo effect is included. The solid lines represent the intensity for $M_{\rm PBH} = 10^3~M_\odot$, $10^2~M_\odot$ and $10~M_\odot$ from top to bottoms. The dotted-dashed line represents the intensity of the diffuse free-free component measured by Planck.}
    \label{fig:intensity_DM}
\end{figure}

Next we take into account the DM halo effect.
The DM halo effect can increase the accretion rate. Therefore the intensity is strongly enhanced. We plot the intenisty with the DM halo effect as a function of the PBH fraction in Fig.~\ref{fig:intensity_DM}.
The DM halo effect also enhances high energy photons which can ionize and heat the background IGM. 
As $f_{\rm PBH}$ increases, the heating and ionization due to PBHs cannot be negligible. 
The heating induce high gas pressure and, accordingly, suppresses the accretion rate. 
Therefore, because of this heating, the intensity decreases when $f_{\rm PBH}$ becomes larger than $f_{\rm PBH} \sim 0.1$ in these mass range.

\begin{figure}
  \centerline{
  \includegraphics[scale=0.6]{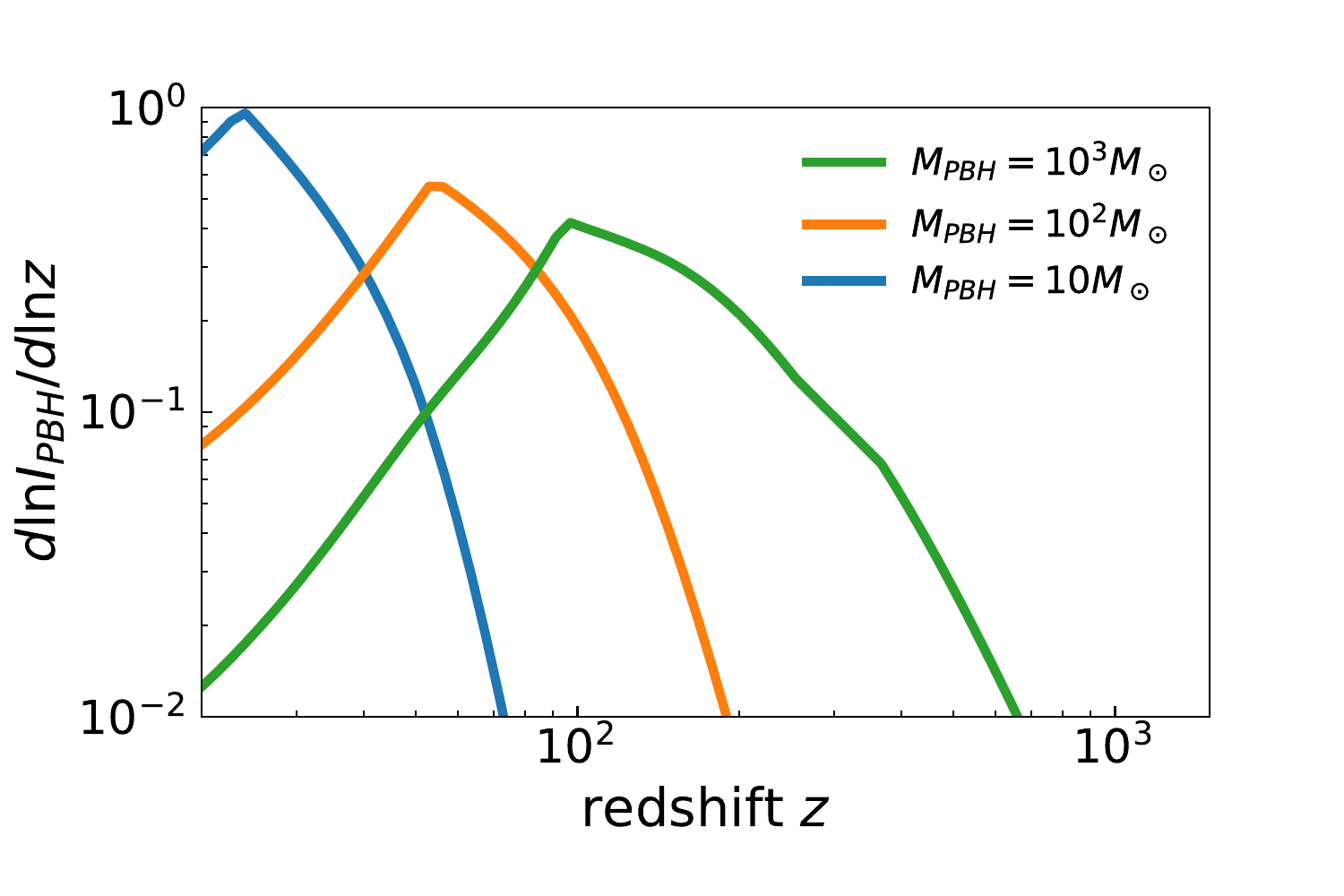}
  }
  \caption{Intensity of the diffuse free-free background radiation from PBHs. Here the DM halo effect is not included. The solid lines represent the intensity for $M_{\rm PBH} = 10^3~M_\odot$, $10^2~M_\odot$ and $10~M_\odot$ from top to bottoms.}
    \label{fig:redshift_DM}
\end{figure}

Fig.~\ref{fig:redshift_DM} shows the redshift distribution of the intensity with the DM halo effect. 
Compared with the case without the DM halo effect in Fig.~\ref{fig:redshift},
the contribution from low redshifts is significant in the case with the DM halo effect.
The DM halo mass grows proportionally to $(1+z)^{-1}$ and the total mass gravitationally attracting the gas becomes 100 times larger than the PBH mass.
As a result, the enhancement of the accretion rate appears in lower redshifts.
This is the reason why the intensity is amplified in low redshifts in the case with the DM halo effect.
However, as the DM mass increases, the heating effect cannot be negligible. As mentioned above the heating suppresses the accretion rate.
As the PBH mass increases, the heating effect appears in higher redshifts. 
As a result, the suppression of the accretion rate also arises in high redshifts for large PBH mass.
Because of this suppression, the redshift contribution has a peak shape and the location of the peak shifts toward low redshifts when the PBH mass decreases.

In the analysis for the foreground radiation in the CMB frequency range with the Planck data,
Ref.~\cite{2016A&A...594A..10P} identified the free-free emission component as
$I _{\rm ff} \approx 2000 ~\rm Jy/str$~($T_{\rm b,ff}  \approx 35~(\nu_{\rm obs} /44~{\rm GHz})^{-2}~\rm \mu K$ in terms of the brigntness temperature, $T_{\rm b}$). 
This is consistent with the results in ARCADE~2~\cite{2011ApJ...734....5F}.
Comparing with the observed intensity,
we find out that our results are much below it. 

Most of the observed free-free signals is can be Galactic origin
This is because, in the Planck analysis, the observed free-free emission is strongly anisotropic~\cite{2016A&A...594A..10P} and the strong signals come from the galactic disk region. The signals become much lower in high galactic latitude. Using the Planck free-free emission map, Ref.~\cite{abeetal2021} evaluated that 
the free-free emission in high galactic latitude~$|b| >60$, $I_{\rm ff} \sim 60~\rm Jy/str$ in the CMB frequency range. When we apply this value, we obtain the constraint on the PBH fraction for $M>10^3~M_\odot$.
However the obtained limit is already is ruled out by the constraint from the CMB anisotropy.

\subsection{Free-Free emission at the radio frequency}

The free-free emission has a flat frequency spectrum. 
Therefore, 
although the spectrum is much lower than the CMB blackbody spectrum in the CMB frequency range, it might contribute to the cosmic radio background with the some fraction.
Depending on the fraction,
there is possibility that 
the free-free emission from PBHs in high redshifts
can affect the absorption feature of the global 21-cm signal in the dark ages and cosmic dawn~\cite{2018ApJ...858L..17F}.

In order to study the impact on the global 21-cm signals,
we calculate the brightness temperature of the free-free emission from PBHs
at $1420~\rm{MHz}$ in the 
redshift $z = 15$.
In such lower frequency, 
we cannot neglect the absorption effect and
the optically thin approximation is not valid.
We calculate the intensity and obtain the fittig form for $f_{\rm PBH} <0.1$ as
\begin{align}
\frac{T_{\rm b, PBH}}{T_{\gamma}} \approx 10^{-3} \left(\frac{f_{\rm PBH}}{10^{-2}} \right)^{1/2}
\left(\frac{M_{\rm PBH}}{100 ~M_{\odot} }\right)^{1.3},
\end{align}
at $1420~\rm{MHz}$ in the 
redshift $z = 15$.

Recently, the EDGES collaboration has reported the larger absorption signals of redshifted 21-cm lines between redshift $z = 20$ and $z = 15$~\cite{2018Natur.555...67B}. The observed absorption is two times stronger than the maximum value in the predictions with the standard cosmological model.
One of possibilities to explain such large absorption is the enhancement of the background radiation at $1420~\rm{MHz}$ in the 
redshift $z = 15$.
Although 
the required brightness temperature of the free-free emission 
depends on the gas temperature and the strength of the Lyman-$\alpha$ background radiation,
the required brightness temperature is twice the CMB temperature
in the simple case, where gas is adiabatically cooled and there exists strong Lyman-$\alpha$ background from stars.

Unfortunately our results shows that the free-free emission intensity from PBHs cannot reach the level to explain the EDGES result.
The impact on the global signal is less than the percent level 
for PBH with allowed $f_{\rm PBH}$.

% Ref.~\cite{2018PhRvD..98b3503H} has already studied the constraint on 
% the PBH abundance based on the EDGES report. However, they do not consider the free-free emission as the background radiation.
% Our results tell us that the free-free emission cannot be negligible. 
% To predict the global 21-cm signal
% with the free-free emission from PBHs, it is also required to evaluate the impact of the free-free emission on thermal evolution of the background baryon gas.
% However, it is beyond the scope of this paper. We address this issue and the constraint on the PBH in the next paper.

\section{conclusion}

PBHs accrete the baryon gas
by their gravity. The accreting gas is heated up
and ionized due to the compression during the accretion.
Such hot ionized gas can emit free-free emissions, and
the sum of these emissions contributes to the diffuse free-free background radiation. 
In this paper, we have calculated the intensity of the free-free background radiation and obtained the relation between the intensity and the PBH abundance.
We have found that considering the growth of the DM halo around a PBH enhances the free-free emission strongly. Because of the DM growth, the emission contribution from lower redshifts dominates the one from higher redshifts.

Comparing with the diffuse free-free component measured in the CMB frequency,
the free-free emission is much lower than the observed signals.
Therefore, it is difficult to obtain the constraint from the free-free emission signals.

The observed free-free emission is highly anisotropic. 
In particular, the strong emission comes from the Galactic disk regions.
It is difficult to identify the free-free emission from the cosmological origin (i.e. from the outside of Galaxies). 
However the upcoming SKA observation might be able to detect the extra-Galactic free-free emission signal.
although it is required that one perform the careful study with other low-frequency radio point sources~\cite{2004ApJ...606L...5C},
the further understand of the cosmological free-free emission signal
might be helpful to obtain the PBH abundance limit.

There is the possibility that the stellar mass and massive PBHs form the accretion disk~\cite{2017PhRvD..96h3524P}.
The radiation from the disk can be stronger than
in the spherical case as in this paper.
However, when the radiation is strong, it induces the photoionization and heats up the gas in the cosmological scales~\cite{2013MNRAS.435.3001T,2019PhRvD..99j3519A}. To evaluate these effects, a numerical simulation is required.
We also leave them for our future work.

\if0
\begin{align}
\Delta M &= \dot M t_{\rm cos} = \frac{\dot M c^2} {L_{edd}} t_{\rm cos} \frac{L_{edd}}{c^2}\\
\frac{\Delta M}{M}& = \frac{\dot M c^2} {L_{edd}} t_{\rm cos} \frac{L_{edd}}{M c^2}.\\
{\rm For~}&M=10^4 M_\odot\\
\frac{\Delta M}{M}& = 0.1 \times  (1~{\rm Gyr}) \frac{(M/M_\odot)\times 10^{31} ~\rm W}{(M/M_\odot) \times 10^{47} \rm Joule}\\
&=0.3
\end{align}
\fi

\acknowledgements
This work is supported by Japan Society for the Promotion of Science~(JSPS) KAKENHI GrantsNo.~JP21K03533 (H.T) and
No.~JP20J22260 (K.T.A),
and supported by JSPS Overseas Research Fellowships (T.M.).

\bibliography{PBH_freefree}% Produces the bibliography via BibTeX.

\providecommand{\noopsort}[1]{}\providecommand{\singleletter}[1]{#1}%
\begin{thebibliography}{36}%
\makeatletter
\providecommand \@ifxundefined [1]{%
 \@ifx{#1\undefined}
}%
\providecommand \@ifnum [1]{%
 \ifnum #1\expandafter \@firstoftwo
 \else \expandafter \@secondoftwo
 \fi
}%
\providecommand \@ifx [1]{%
 \ifx #1\expandafter \@firstoftwo
 \else \expandafter \@secondoftwo
 \fi
}%
\providecommand \natexlab [1]{#1}%
\providecommand \enquote  [1]{``#1''}%
\providecommand \bibnamefont  [1]{#1}%
\providecommand \bibfnamefont [1]{#1}%
\providecommand \citenamefont [1]{#1}%
\providecommand \href@noop [0]{\@secondoftwo}%
\providecommand \href [0]{\begingroup \@sanitize@url \@href}%
\providecommand \@href[1]{\@@startlink{#1}\@@href}%
\providecommand \@@href[1]{\endgroup#1\@@endlink}%
\providecommand \@sanitize@url [0]{\catcode `\\12\catcode `\$12\catcode
  `\&12\catcode `\#12\catcode `\^12\catcode `\_12\catcode `\%12\relax}%
\providecommand \@@startlink[1]{}%
\providecommand \@@endlink[0]{}%
\providecommand \url  [0]{\begingroup\@sanitize@url \@url }%
\providecommand \@url [1]{\endgroup\@href {#1}{\urlprefix }}%
\providecommand \urlprefix  [0]{URL }%
\providecommand \Eprint [0]{\href }%
\providecommand \doibase [0]{https://doi.org/}%
\providecommand \selectlanguage [0]{\@gobble}%
\providecommand \bibinfo  [0]{\@secondoftwo}%
\providecommand \bibfield  [0]{\@secondoftwo}%
\providecommand \translation [1]{[#1]}%
\providecommand \BibitemOpen [0]{}%
\providecommand \bibitemStop [0]{}%
\providecommand \bibitemNoStop [0]{.\EOS\space}%
\providecommand \EOS [0]{\spacefactor3000\relax}%
\providecommand \BibitemShut  [1]{\csname bibitem#1\endcsname}%
\let\auto@bib@innerbib\@empty
%</preamble>
\bibitem [{\citenamefont {{Clowe}}\ \emph {et~al.}(2006)\citenamefont
  {{Clowe}}, \citenamefont {{Brada{\v{c}}}}, \citenamefont {{Gonzalez}},
  \citenamefont {{Markevitch}}, \citenamefont {{Randall}}, \citenamefont
  {{Jones}},\ and\ \citenamefont {{Zaritsky}}}]{2006ApJ...648L.109C}%
  \BibitemOpen
  \bibfield  {author} {\bibinfo {author} {\bibfnamefont {D.}~\bibnamefont
  {{Clowe}}}, \bibinfo {author} {\bibfnamefont {M.}~\bibnamefont
  {{Brada{\v{c}}}}}, \bibinfo {author} {\bibfnamefont {A.~H.}\ \bibnamefont
  {{Gonzalez}}}, \bibinfo {author} {\bibfnamefont {M.}~\bibnamefont
  {{Markevitch}}}, \bibinfo {author} {\bibfnamefont {S.~W.}\ \bibnamefont
  {{Randall}}}, \bibinfo {author} {\bibfnamefont {C.}~\bibnamefont {{Jones}}},\
  and\ \bibinfo {author} {\bibfnamefont {D.}~\bibnamefont {{Zaritsky}}},\
  }\bibfield  {title} {\bibinfo {title} {{A Direct Empirical Proof of the
  Existence of Dark Matter}},\ }\href {https://doi.org/10.1086/508162}
  {\bibfield  {journal} {\bibinfo  {journal} {\apjl}\ }\textbf {\bibinfo
  {volume} {648}},\ \bibinfo {pages} {L109} (\bibinfo {year} {2006})},\ \Eprint
  {https://arxiv.org/abs/astro-ph/0608407} {arXiv:astro-ph/0608407 [astro-ph]}
  \BibitemShut {NoStop}%
\bibitem [{\citenamefont {{Planck Collaboration}}(2020)}]{2020A&A...641A...6P}%
  \BibitemOpen
  \bibfield  {author} {\bibinfo {author} {\bibnamefont {{Planck
  Collaboration}}},\ }\bibfield  {title} {\bibinfo {title} {{Planck 2018
  results. VI. Cosmological parameters}},\ }\href
  {https://doi.org/10.1051/0004-6361/201833910} {\bibfield  {journal} {\bibinfo
   {journal} {\aap}\ }\textbf {\bibinfo {volume} {641}},\ \bibinfo {eid} {A6}
  (\bibinfo {year} {2020})},\ \Eprint {https://arxiv.org/abs/1807.06209}
  {arXiv:1807.06209 [astro-ph.CO]} \BibitemShut {NoStop}%
\bibitem [{\citenamefont {{Jungman}}\ \emph {et~al.}(1996)\citenamefont
  {{Jungman}}, \citenamefont {{Kamionkowski}},\ and\ \citenamefont
  {{Griest}}}]{1996PhR...267..195J}%
  \BibitemOpen
  \bibfield  {author} {\bibinfo {author} {\bibfnamefont {G.}~\bibnamefont
  {{Jungman}}}, \bibinfo {author} {\bibfnamefont {M.}~\bibnamefont
  {{Kamionkowski}}},\ and\ \bibinfo {author} {\bibfnamefont {K.}~\bibnamefont
  {{Griest}}},\ }\bibfield  {title} {\bibinfo {title} {{Supersymmetric dark
  matter}},\ }\href {https://doi.org/10.1016/0370-1573(95)00058-5} {\bibfield
  {journal} {\bibinfo  {journal} {\physrep}\ }\textbf {\bibinfo {volume}
  {267}},\ \bibinfo {pages} {195} (\bibinfo {year} {1996})},\ \Eprint
  {https://arxiv.org/abs/hep-ph/9506380} {arXiv:hep-ph/9506380 [hep-ph]}
  \BibitemShut {NoStop}%
\bibitem [{\citenamefont {{Bertone}}\ \emph {et~al.}(2005)\citenamefont
  {{Bertone}}, \citenamefont {{Hooper}},\ and\ \citenamefont
  {{Silk}}}]{2005PhR...405..279B}%
  \BibitemOpen
  \bibfield  {author} {\bibinfo {author} {\bibfnamefont {G.}~\bibnamefont
  {{Bertone}}}, \bibinfo {author} {\bibfnamefont {D.}~\bibnamefont
  {{Hooper}}},\ and\ \bibinfo {author} {\bibfnamefont {J.}~\bibnamefont
  {{Silk}}},\ }\bibfield  {title} {\bibinfo {title} {{Particle dark matter:
  evidence, candidates and constraints}},\ }\href
  {https://doi.org/10.1016/j.physrep.2004.08.031} {\bibfield  {journal}
  {\bibinfo  {journal} {\physrep}\ }\textbf {\bibinfo {volume} {405}},\
  \bibinfo {pages} {279} (\bibinfo {year} {2005})},\ \Eprint
  {https://arxiv.org/abs/hep-ph/0404175} {arXiv:hep-ph/0404175 [hep-ph]}
  \BibitemShut {NoStop}%
\bibitem [{\citenamefont {{Zel'dovich}}\ and\ \citenamefont
  {{Novikov}}(1967)}]{1967SvA....10..602Z}%
  \BibitemOpen
  \bibfield  {author} {\bibinfo {author} {\bibfnamefont {Y.~B.}\ \bibnamefont
  {{Zel'dovich}}}\ and\ \bibinfo {author} {\bibfnamefont {I.~D.}\ \bibnamefont
  {{Novikov}}},\ }\bibfield  {title} {\bibinfo {title} {{The Hypothesis of
  Cores Retarded during Expansion and the Hot Cosmological Model}},\
  }\href@noop {} {\bibfield  {journal} {\bibinfo  {journal} {\sovast}\ }\textbf
  {\bibinfo {volume} {10}},\ \bibinfo {pages} {602} (\bibinfo {year}
  {1967})}\BibitemShut {NoStop}%
\bibitem [{\citenamefont {{Hawking}}(1971)}]{1971MNRAS.152...75H}%
  \BibitemOpen
  \bibfield  {author} {\bibinfo {author} {\bibfnamefont {S.}~\bibnamefont
  {{Hawking}}},\ }\bibfield  {title} {\bibinfo {title} {{Gravitationally
  collapsed objects of very low mass}},\ }\href
  {https://doi.org/10.1093/mnras/152.1.75} {\bibfield  {journal} {\bibinfo
  {journal} {\mnras}\ }\textbf {\bibinfo {volume} {152}},\ \bibinfo {pages}
  {75} (\bibinfo {year} {1971})}\BibitemShut {NoStop}%
\bibitem [{\citenamefont {{Carr}}\ and\ \citenamefont
  {{Hawking}}(1974)}]{1974MNRAS.168..399C}%
  \BibitemOpen
  \bibfield  {author} {\bibinfo {author} {\bibfnamefont {B.~J.}\ \bibnamefont
  {{Carr}}}\ and\ \bibinfo {author} {\bibfnamefont {S.~W.}\ \bibnamefont
  {{Hawking}}},\ }\bibfield  {title} {\bibinfo {title} {{Black holes in the
  early Universe}},\ }\href {https://doi.org/10.1093/mnras/168.2.399}
  {\bibfield  {journal} {\bibinfo  {journal} {\mnras}\ }\textbf {\bibinfo
  {volume} {168}},\ \bibinfo {pages} {399} (\bibinfo {year}
  {1974})}\BibitemShut {NoStop}%
\bibitem [{\citenamefont {{Chapline}}(1975)}]{1975Natur.253..251C}%
  \BibitemOpen
  \bibfield  {author} {\bibinfo {author} {\bibfnamefont {G.~F.}\ \bibnamefont
  {{Chapline}}},\ }\bibfield  {title} {\bibinfo {title} {{Cosmological effects
  of primordial black holes}},\ }\href {https://doi.org/10.1038/253251a0}
  {\bibfield  {journal} {\bibinfo  {journal} {\nat}\ }\textbf {\bibinfo
  {volume} {253}},\ \bibinfo {pages} {251} (\bibinfo {year}
  {1975})}\BibitemShut {NoStop}%
\bibitem [{\citenamefont {{Bird}}\ \emph {et~al.}(2016)\citenamefont {{Bird}},
  \citenamefont {{Cholis}}, \citenamefont {{Mu{\~n}oz}}, \citenamefont
  {{Ali-Ha{\"\i}moud}}, \citenamefont {{Kamionkowski}}, \citenamefont
  {{Kovetz}}, \citenamefont {{Raccanelli}},\ and\ \citenamefont
  {{Riess}}}]{2016PhRvL.116t1301B}%
  \BibitemOpen
  \bibfield  {author} {\bibinfo {author} {\bibfnamefont {S.}~\bibnamefont
  {{Bird}}}, \bibinfo {author} {\bibfnamefont {I.}~\bibnamefont {{Cholis}}},
  \bibinfo {author} {\bibfnamefont {J.~B.}\ \bibnamefont {{Mu{\~n}oz}}},
  \bibinfo {author} {\bibfnamefont {Y.}~\bibnamefont {{Ali-Ha{\"\i}moud}}},
  \bibinfo {author} {\bibfnamefont {M.}~\bibnamefont {{Kamionkowski}}},
  \bibinfo {author} {\bibfnamefont {E.~D.}\ \bibnamefont {{Kovetz}}}, \bibinfo
  {author} {\bibfnamefont {A.}~\bibnamefont {{Raccanelli}}},\ and\ \bibinfo
  {author} {\bibfnamefont {A.~G.}\ \bibnamefont {{Riess}}},\ }\bibfield
  {title} {\bibinfo {title} {{Did LIGO Detect Dark Matter?}},\ }\href
  {https://doi.org/10.1103/PhysRevLett.116.201301} {\bibfield  {journal}
  {\bibinfo  {journal} {\prl}\ }\textbf {\bibinfo {volume} {116}},\ \bibinfo
  {eid} {201301} (\bibinfo {year} {2016})},\ \Eprint
  {https://arxiv.org/abs/1603.00464} {arXiv:1603.00464 [astro-ph.CO]}
  \BibitemShut {NoStop}%
\bibitem [{\citenamefont {{Carr}}\ and\ \citenamefont
  {{K{\"u}hnel}}(2020)}]{2020ARNPS..70..355C}%
  \BibitemOpen
  \bibfield  {author} {\bibinfo {author} {\bibfnamefont {B.}~\bibnamefont
  {{Carr}}}\ and\ \bibinfo {author} {\bibfnamefont {F.}~\bibnamefont
  {{K{\"u}hnel}}},\ }\bibfield  {title} {\bibinfo {title} {{Primordial Black
  Holes as Dark Matter: Recent Developments}},\ }\href
  {https://doi.org/10.1146/annurev-nucl-050520-125911} {\bibfield  {journal}
  {\bibinfo  {journal} {Annual Review of Nuclear and Particle Science}\
  }\textbf {\bibinfo {volume} {70}},\ \bibinfo {pages} {355} (\bibinfo {year}
  {2020})},\ \Eprint {https://arxiv.org/abs/2006.02838} {arXiv:2006.02838
  [astro-ph.CO]} \BibitemShut {NoStop}%
\bibitem [{\citenamefont {{Green}}\ and\ \citenamefont
  {{Kavanagh}}(2021)}]{2021JPhG...48d3001G}%
  \BibitemOpen
  \bibfield  {author} {\bibinfo {author} {\bibfnamefont {A.~M.}\ \bibnamefont
  {{Green}}}\ and\ \bibinfo {author} {\bibfnamefont {B.~J.}\ \bibnamefont
  {{Kavanagh}}},\ }\bibfield  {title} {\bibinfo {title} {{Primordial black
  holes as a dark matter candidate}},\ }\href
  {https://doi.org/10.1088/1361-6471/abc534} {\bibfield  {journal} {\bibinfo
  {journal} {Journal of Physics G Nuclear Physics}\ }\textbf {\bibinfo {volume}
  {48}},\ \bibinfo {eid} {043001} (\bibinfo {year} {2021})},\ \Eprint
  {https://arxiv.org/abs/2007.10722} {arXiv:2007.10722 [astro-ph.CO]}
  \BibitemShut {NoStop}%
\bibitem [{\citenamefont {{Ricotti}}\ \emph {et~al.}(2008)\citenamefont
  {{Ricotti}}, \citenamefont {{Ostriker}},\ and\ \citenamefont
  {{Mack}}}]{2008ApJ...680..829R}%
  \BibitemOpen
  \bibfield  {author} {\bibinfo {author} {\bibfnamefont {M.}~\bibnamefont
  {{Ricotti}}}, \bibinfo {author} {\bibfnamefont {J.~P.}\ \bibnamefont
  {{Ostriker}}},\ and\ \bibinfo {author} {\bibfnamefont {K.~J.}\ \bibnamefont
  {{Mack}}},\ }\bibfield  {title} {\bibinfo {title} {{Effect of Primordial
  Black Holes on the Cosmic Microwave Background and Cosmological Parameter
  Estimates}},\ }\href {https://doi.org/10.1086/587831} {\bibfield  {journal}
  {\bibinfo  {journal} {\apj}\ }\textbf {\bibinfo {volume} {680}},\ \bibinfo
  {pages} {829} (\bibinfo {year} {2008})},\ \Eprint
  {https://arxiv.org/abs/0709.0524} {arXiv:0709.0524 [astro-ph]} \BibitemShut
  {NoStop}%
\bibitem [{\citenamefont {{Ali-Ha{\"\i}moud}}\ and\ \citenamefont
  {{Kamionkowski}}(2017)}]{2017PhRvD..95d3534A}%
  \BibitemOpen
  \bibfield  {author} {\bibinfo {author} {\bibfnamefont {Y.}~\bibnamefont
  {{Ali-Ha{\"\i}moud}}}\ and\ \bibinfo {author} {\bibfnamefont
  {M.}~\bibnamefont {{Kamionkowski}}},\ }\bibfield  {title} {\bibinfo {title}
  {{Cosmic microwave background limits on accreting primordial black holes}},\
  }\href {https://doi.org/10.1103/PhysRevD.95.043534} {\bibfield  {journal}
  {\bibinfo  {journal} {\prd}\ }\textbf {\bibinfo {volume} {95}},\ \bibinfo
  {eid} {043534} (\bibinfo {year} {2017})},\ \Eprint
  {https://arxiv.org/abs/1612.05644} {arXiv:1612.05644 [astro-ph.CO]}
  \BibitemShut {NoStop}%
\bibitem [{\citenamefont {{Poulin}}\ \emph {et~al.}(2017)\citenamefont
  {{Poulin}}, \citenamefont {{Serpico}}, \citenamefont {{Calore}},
  \citenamefont {{Clesse}},\ and\ \citenamefont
  {{Kohri}}}]{2017PhRvD..96h3524P}%
  \BibitemOpen
  \bibfield  {author} {\bibinfo {author} {\bibfnamefont {V.}~\bibnamefont
  {{Poulin}}}, \bibinfo {author} {\bibfnamefont {P.~D.}\ \bibnamefont
  {{Serpico}}}, \bibinfo {author} {\bibfnamefont {F.}~\bibnamefont {{Calore}}},
  \bibinfo {author} {\bibfnamefont {S.}~\bibnamefont {{Clesse}}},\ and\
  \bibinfo {author} {\bibfnamefont {K.}~\bibnamefont {{Kohri}}},\ }\bibfield
  {title} {\bibinfo {title} {{CMB bounds on disk-accreting massive primordial
  black holes}},\ }\href {https://doi.org/10.1103/PhysRevD.96.083524}
  {\bibfield  {journal} {\bibinfo  {journal} {\prd}\ }\textbf {\bibinfo
  {volume} {96}},\ \bibinfo {eid} {083524} (\bibinfo {year} {2017})},\ \Eprint
  {https://arxiv.org/abs/1707.04206} {arXiv:1707.04206 [astro-ph.CO]}
  \BibitemShut {NoStop}%
\bibitem [{\citenamefont {{Serpico}}\ \emph {et~al.}(2020)\citenamefont
  {{Serpico}}, \citenamefont {{Poulin}}, \citenamefont {{Inman}},\ and\
  \citenamefont {{Kohri}}}]{2020PhRvR...2b3204S}%
  \BibitemOpen
  \bibfield  {author} {\bibinfo {author} {\bibfnamefont {P.~D.}\ \bibnamefont
  {{Serpico}}}, \bibinfo {author} {\bibfnamefont {V.}~\bibnamefont {{Poulin}}},
  \bibinfo {author} {\bibfnamefont {D.}~\bibnamefont {{Inman}}},\ and\ \bibinfo
  {author} {\bibfnamefont {K.}~\bibnamefont {{Kohri}}},\ }\bibfield  {title}
  {\bibinfo {title} {{Cosmic microwave background bounds on primordial black
  holes including dark matter halo accretion}},\ }\href
  {https://doi.org/10.1103/PhysRevResearch.2.023204} {\bibfield  {journal}
  {\bibinfo  {journal} {Physical Review Research}\ }\textbf {\bibinfo {volume}
  {2}},\ \bibinfo {eid} {023204} (\bibinfo {year} {2020})},\ \Eprint
  {https://arxiv.org/abs/2002.10771} {arXiv:2002.10771 [astro-ph.CO]}
  \BibitemShut {NoStop}%
\bibitem [{\citenamefont {{Abe}}\ \emph {et~al.}(2019)\citenamefont {{Abe}},
  \citenamefont {{Tashiro}},\ and\ \citenamefont
  {{Tanaka}}}]{2019PhRvD..99j3519A}%
  \BibitemOpen
  \bibfield  {author} {\bibinfo {author} {\bibfnamefont {K.~T.}\ \bibnamefont
  {{Abe}}}, \bibinfo {author} {\bibfnamefont {H.}~\bibnamefont {{Tashiro}}},\
  and\ \bibinfo {author} {\bibfnamefont {T.}~\bibnamefont {{Tanaka}}},\
  }\bibfield  {title} {\bibinfo {title} {{Thermal Sunyaev-Zel'dovich anisotropy
  due to primordial black holes}},\ }\href
  {https://doi.org/10.1103/PhysRevD.99.103519} {\bibfield  {journal} {\bibinfo
  {journal} {\prd}\ }\textbf {\bibinfo {volume} {99}},\ \bibinfo {eid} {103519}
  (\bibinfo {year} {2019})},\ \Eprint {https://arxiv.org/abs/1901.06809}
  {arXiv:1901.06809 [astro-ph.CO]} \BibitemShut {NoStop}%
\bibitem [{\citenamefont {{Tashiro}}\ and\ \citenamefont
  {{Sugiyama}}(2013)}]{2013MNRAS.435.3001T}%
  \BibitemOpen
  \bibfield  {author} {\bibinfo {author} {\bibfnamefont {H.}~\bibnamefont
  {{Tashiro}}}\ and\ \bibinfo {author} {\bibfnamefont {N.}~\bibnamefont
  {{Sugiyama}}},\ }\bibfield  {title} {\bibinfo {title} {{The effect of
  primordial black holes on 21-cm fluctuations}},\ }\href
  {https://doi.org/10.1093/mnras/stt1493} {\bibfield  {journal} {\bibinfo
  {journal} {\mnras}\ }\textbf {\bibinfo {volume} {435}},\ \bibinfo {pages}
  {3001} (\bibinfo {year} {2013})},\ \Eprint {https://arxiv.org/abs/1207.6405}
  {arXiv:1207.6405 [astro-ph.CO]} \BibitemShut {NoStop}%
\bibitem [{\citenamefont {{Gong}}\ and\ \citenamefont
  {{Kitajima}}(2017)}]{2017JCAP...08..017G}%
  \BibitemOpen
  \bibfield  {author} {\bibinfo {author} {\bibfnamefont {J.-O.}\ \bibnamefont
  {{Gong}}}\ and\ \bibinfo {author} {\bibfnamefont {N.}~\bibnamefont
  {{Kitajima}}},\ }\bibfield  {title} {\bibinfo {title} {{Small-scale structure
  and 21cm fluctuations by primordial black holes}},\ }\href
  {https://doi.org/10.1088/1475-7516/2017/08/017} {\bibfield  {journal}
  {\bibinfo  {journal} {\jcap}\ }\textbf {\bibinfo {volume} {2017}},\ \bibinfo
  {eid} {017} (\bibinfo {year} {2017})},\ \Eprint
  {https://arxiv.org/abs/1704.04132} {arXiv:1704.04132 [astro-ph.CO]}
  \BibitemShut {NoStop}%
\bibitem [{\citenamefont {{Hektor}}\ \emph {et~al.}(2018)\citenamefont
  {{Hektor}}, \citenamefont {{H{\"u}tsi}}, \citenamefont {{Marzola}},
  \citenamefont {{Raidal}}, \citenamefont {{Vaskonen}},\ and\ \citenamefont
  {{Veerm{\"a}e}}}]{2018PhRvD..98b3503H}%
  \BibitemOpen
  \bibfield  {author} {\bibinfo {author} {\bibfnamefont {A.}~\bibnamefont
  {{Hektor}}}, \bibinfo {author} {\bibfnamefont {G.}~\bibnamefont
  {{H{\"u}tsi}}}, \bibinfo {author} {\bibfnamefont {L.}~\bibnamefont
  {{Marzola}}}, \bibinfo {author} {\bibfnamefont {M.}~\bibnamefont {{Raidal}}},
  \bibinfo {author} {\bibfnamefont {V.}~\bibnamefont {{Vaskonen}}},\ and\
  \bibinfo {author} {\bibfnamefont {H.}~\bibnamefont {{Veerm{\"a}e}}},\
  }\bibfield  {title} {\bibinfo {title} {{Constraining primordial black holes
  with the EDGES 21-cm absorption signal}},\ }\href
  {https://doi.org/10.1103/PhysRevD.98.023503} {\bibfield  {journal} {\bibinfo
  {journal} {\prd}\ }\textbf {\bibinfo {volume} {98}},\ \bibinfo {eid} {023503}
  (\bibinfo {year} {2018})},\ \Eprint {https://arxiv.org/abs/1803.09697}
  {arXiv:1803.09697 [astro-ph.CO]} \BibitemShut {NoStop}%
\bibitem [{\citenamefont {{Mena}}\ \emph {et~al.}(2019)\citenamefont {{Mena}},
  \citenamefont {{Palomares-Ruiz}}, \citenamefont {{Villanueva-Domingo}},\ and\
  \citenamefont {{Witte}}}]{2019PhRvD.100d3540M}%
  \BibitemOpen
  \bibfield  {author} {\bibinfo {author} {\bibfnamefont {O.}~\bibnamefont
  {{Mena}}}, \bibinfo {author} {\bibfnamefont {S.}~\bibnamefont
  {{Palomares-Ruiz}}}, \bibinfo {author} {\bibfnamefont {P.}~\bibnamefont
  {{Villanueva-Domingo}}},\ and\ \bibinfo {author} {\bibfnamefont {S.~J.}\
  \bibnamefont {{Witte}}},\ }\bibfield  {title} {\bibinfo {title}
  {{Constraining the primordial black hole abundance with 21-cm cosmology}},\
  }\href {https://doi.org/10.1103/PhysRevD.100.043540} {\bibfield  {journal}
  {\bibinfo  {journal} {\prd}\ }\textbf {\bibinfo {volume} {100}},\ \bibinfo
  {eid} {043540} (\bibinfo {year} {2019})},\ \Eprint
  {https://arxiv.org/abs/1906.07735} {arXiv:1906.07735 [astro-ph.CO]}
  \BibitemShut {NoStop}%
\bibitem [{\citenamefont {{Seiffert}}\ \emph {et~al.}(2011)\citenamefont
  {{Seiffert}}, \citenamefont {{Fixsen}}, \citenamefont {{Kogut}},
  \citenamefont {{Levin}}, \citenamefont {{Limon}}, \citenamefont {{Lubin}},
  \citenamefont {{Mirel}}, \citenamefont {{Singal}}, \citenamefont {{Villela}},
  \citenamefont {{Wollack}},\ and\ \citenamefont
  {{Wuensche}}}]{2011ApJ...734....6S}%
  \BibitemOpen
  \bibfield  {author} {\bibinfo {author} {\bibfnamefont {M.}~\bibnamefont
  {{Seiffert}}}, \bibinfo {author} {\bibfnamefont {D.~J.}\ \bibnamefont
  {{Fixsen}}}, \bibinfo {author} {\bibfnamefont {A.}~\bibnamefont {{Kogut}}},
  \bibinfo {author} {\bibfnamefont {S.~M.}\ \bibnamefont {{Levin}}}, \bibinfo
  {author} {\bibfnamefont {M.}~\bibnamefont {{Limon}}}, \bibinfo {author}
  {\bibfnamefont {P.~M.}\ \bibnamefont {{Lubin}}}, \bibinfo {author}
  {\bibfnamefont {P.}~\bibnamefont {{Mirel}}}, \bibinfo {author} {\bibfnamefont
  {J.}~\bibnamefont {{Singal}}}, \bibinfo {author} {\bibfnamefont
  {T.}~\bibnamefont {{Villela}}}, \bibinfo {author} {\bibfnamefont
  {E.}~\bibnamefont {{Wollack}}},\ and\ \bibinfo {author} {\bibfnamefont
  {C.~A.}\ \bibnamefont {{Wuensche}}},\ }\bibfield  {title} {\bibinfo {title}
  {{Interpretation of the ARCADE 2 Absolute Sky Brightness Measurement}},\
  }\href {https://doi.org/10.1088/0004-637X/734/1/6} {\bibfield  {journal}
  {\bibinfo  {journal} {\apj}\ }\textbf {\bibinfo {volume} {734}},\ \bibinfo
  {eid} {6} (\bibinfo {year} {2011})}\BibitemShut {NoStop}%
\bibitem [{\citenamefont {{Planck Collaboration}}(2016)}]{2016A&A...594A..10P}%
  \BibitemOpen
  \bibfield  {author} {\bibinfo {author} {\bibnamefont {{Planck
  Collaboration}}},\ }\bibfield  {title} {\bibinfo {title} {{Planck 2015
  results. X. Diffuse component separation: Foreground maps}},\ }\href
  {https://doi.org/10.1051/0004-6361/201525967} {\bibfield  {journal} {\bibinfo
   {journal} {\aap}\ }\textbf {\bibinfo {volume} {594}},\ \bibinfo {eid} {A10}
  (\bibinfo {year} {2016})},\ \Eprint {https://arxiv.org/abs/1502.01588}
  {arXiv:1502.01588 [astro-ph.CO]} \BibitemShut {NoStop}%
\bibitem [{\citenamefont {{Cooray}}\ and\ \citenamefont
  {{Furlanetto}}(2004)}]{2004ApJ...606L...5C}%
  \BibitemOpen
  \bibfield  {author} {\bibinfo {author} {\bibfnamefont {A.}~\bibnamefont
  {{Cooray}}}\ and\ \bibinfo {author} {\bibfnamefont {S.~R.}\ \bibnamefont
  {{Furlanetto}}},\ }\bibfield  {title} {\bibinfo {title} {{Free-Free Emission
  at Low Radio Frequencies}},\ }\href {https://doi.org/10.1086/421241}
  {\bibfield  {journal} {\bibinfo  {journal} {\apjl}\ }\textbf {\bibinfo
  {volume} {606}},\ \bibinfo {pages} {L5} (\bibinfo {year} {2004})},\ \Eprint
  {https://arxiv.org/abs/astro-ph/0402239} {arXiv:astro-ph/0402239 [astro-ph]}
  \BibitemShut {NoStop}%
\bibitem [{\citenamefont {{Ponente}}\ \emph {et~al.}(2011)\citenamefont
  {{Ponente}}, \citenamefont {{Diego}}, \citenamefont {{Sheth}}, \citenamefont
  {{Burigana}}, \citenamefont {{Knollmann}},\ and\ \citenamefont
  {{Ascasibar}}}]{2011MNRAS.410.2353P}%
  \BibitemOpen
  \bibfield  {author} {\bibinfo {author} {\bibfnamefont {P.~P.}\ \bibnamefont
  {{Ponente}}}, \bibinfo {author} {\bibfnamefont {J.~M.}\ \bibnamefont
  {{Diego}}}, \bibinfo {author} {\bibfnamefont {R.~K.}\ \bibnamefont
  {{Sheth}}}, \bibinfo {author} {\bibfnamefont {C.}~\bibnamefont {{Burigana}}},
  \bibinfo {author} {\bibfnamefont {S.~R.}\ \bibnamefont {{Knollmann}}},\ and\
  \bibinfo {author} {\bibfnamefont {Y.}~\bibnamefont {{Ascasibar}}},\
  }\bibfield  {title} {\bibinfo {title} {{The cosmological free-free signal
  from galaxy groups and clusters}},\ }\href
  {https://doi.org/10.1111/j.1365-2966.2010.17611.x} {\bibfield  {journal}
  {\bibinfo  {journal} {\mnras}\ }\textbf {\bibinfo {volume} {410}},\ \bibinfo
  {pages} {2353} (\bibinfo {year} {2011})},\ \Eprint
  {https://arxiv.org/abs/1006.2243} {arXiv:1006.2243 [astro-ph.CO]}
  \BibitemShut {NoStop}%
\bibitem [{\citenamefont {{Liu}}\ \emph {et~al.}(2019)\citenamefont {{Liu}},
  \citenamefont {{Jaacks}}, \citenamefont {{Finkelstein}},\ and\ \citenamefont
  {{Bromm}}}]{2019MNRAS.486.3617L}%
  \BibitemOpen
  \bibfield  {author} {\bibinfo {author} {\bibfnamefont {B.}~\bibnamefont
  {{Liu}}}, \bibinfo {author} {\bibfnamefont {J.}~\bibnamefont {{Jaacks}}},
  \bibinfo {author} {\bibfnamefont {S.~L.}\ \bibnamefont {{Finkelstein}}},\
  and\ \bibinfo {author} {\bibfnamefont {V.}~\bibnamefont {{Bromm}}},\
  }\bibfield  {title} {\bibinfo {title} {{Global radiation signature from early
  structure formation}},\ }\href {https://doi.org/10.1093/mnras/stz910}
  {\bibfield  {journal} {\bibinfo  {journal} {\mnras}\ }\textbf {\bibinfo
  {volume} {486}},\ \bibinfo {pages} {3617} (\bibinfo {year} {2019})},\ \Eprint
  {https://arxiv.org/abs/1901.08994} {arXiv:1901.08994 [astro-ph.GA]}
  \BibitemShut {NoStop}%
\bibitem [{\citenamefont {{Abe}}\ \emph {et~al.}(2021)\citenamefont {{Abe}},
  \citenamefont {{Minoda}},\ and\ \citenamefont {{Tashiro}}}]{abeetal2021}%
  \BibitemOpen
  \bibfield  {author} {\bibinfo {author} {\bibfnamefont {K.~T.}\ \bibnamefont
  {{Abe}}}, \bibinfo {author} {\bibfnamefont {T.}~\bibnamefont {{Minoda}}},\
  and\ \bibinfo {author} {\bibfnamefont {H.}~\bibnamefont {{Tashiro}}},\
  }\bibfield  {title} {\bibinfo {title} {{Constraint on the early-formed dark
  matter halos using the free-free emission in the Planck foreground
  analysis}},\ }\href@noop {} {\bibfield  {journal} {\bibinfo  {journal} {arXiv
  e-prints}\ ,\ \bibinfo {eid} {arXiv:2108.00621}} (\bibinfo {year} {2021})},\
  \Eprint {https://arxiv.org/abs/2108.00621} {arXiv:2108.00621 [astro-ph.CO]}
  \BibitemShut {NoStop}%
\bibitem [{\citenamefont {{Bondi}}(1952)}]{1952MNRAS.112..195B}%
  \BibitemOpen
  \bibfield  {author} {\bibinfo {author} {\bibfnamefont {H.}~\bibnamefont
  {{Bondi}}},\ }\bibfield  {title} {\bibinfo {title} {{On spherically
  symmetrical accretion}},\ }\href {https://doi.org/10.1093/mnras/112.2.195}
  {\bibfield  {journal} {\bibinfo  {journal} {\mnras}\ }\textbf {\bibinfo
  {volume} {112}},\ \bibinfo {pages} {195} (\bibinfo {year}
  {1952})}\BibitemShut {NoStop}%
\bibitem [{\citenamefont {{Tseliakhovich}}\ and\ \citenamefont
  {{Hirata}}(2010)}]{2010PhRvD..82h3520T}%
  \BibitemOpen
  \bibfield  {author} {\bibinfo {author} {\bibfnamefont {D.}~\bibnamefont
  {{Tseliakhovich}}}\ and\ \bibinfo {author} {\bibfnamefont {C.}~\bibnamefont
  {{Hirata}}},\ }\bibfield  {title} {\bibinfo {title} {{Relative velocity of
  dark matter and baryonic fluids and the formation of the first structures}},\
  }\href {https://doi.org/10.1103/PhysRevD.82.083520} {\bibfield  {journal}
  {\bibinfo  {journal} {\prd}\ }\textbf {\bibinfo {volume} {82}},\ \bibinfo
  {eid} {083520} (\bibinfo {year} {2010})},\ \Eprint
  {https://arxiv.org/abs/1005.2416} {arXiv:1005.2416 [astro-ph.CO]}
  \BibitemShut {NoStop}%
\bibitem [{\citenamefont {{Mack}}\ \emph {et~al.}(2007)\citenamefont {{Mack}},
  \citenamefont {{Ostriker}},\ and\ \citenamefont
  {{Ricotti}}}]{2007ApJ...665.1277M}%
  \BibitemOpen
  \bibfield  {author} {\bibinfo {author} {\bibfnamefont {K.~J.}\ \bibnamefont
  {{Mack}}}, \bibinfo {author} {\bibfnamefont {J.~P.}\ \bibnamefont
  {{Ostriker}}},\ and\ \bibinfo {author} {\bibfnamefont {M.}~\bibnamefont
  {{Ricotti}}},\ }\bibfield  {title} {\bibinfo {title} {{Growth of Structure
  Seeded by Primordial Black Holes}},\ }\href {https://doi.org/10.1086/518998}
  {\bibfield  {journal} {\bibinfo  {journal} {\apj}\ }\textbf {\bibinfo
  {volume} {665}},\ \bibinfo {pages} {1277} (\bibinfo {year} {2007})},\ \Eprint
  {https://arxiv.org/abs/astro-ph/0608642} {arXiv:astro-ph/0608642 [astro-ph]}
  \BibitemShut {NoStop}%
\bibitem [{\citenamefont {{Rybicki}}\ and\ \citenamefont
  {{Lightman}}(1986)}]{Radiative_process_in_Astrophysics}%
  \BibitemOpen
  \bibfield  {author} {\bibinfo {author} {\bibfnamefont {G.~B.}\ \bibnamefont
  {{Rybicki}}}\ and\ \bibinfo {author} {\bibfnamefont {A.~P.}\ \bibnamefont
  {{Lightman}}},\ }\href@noop {} {\bibinfo {title} {{Radiative Processes in
  Astrophysics}}} (\bibinfo {year} {1986})\BibitemShut {NoStop}%
\bibitem [{\citenamefont {{Draine}}(2011)}]{2011piim.book.....D}%
  \BibitemOpen
  \bibfield  {author} {\bibinfo {author} {\bibfnamefont {B.~T.}\ \bibnamefont
  {{Draine}}},\ }\href@noop {} {\bibinfo {title} {{Physics of the Interstellar
  and Intergalactic Medium}}} (\bibinfo {year} {2011})\BibitemShut {NoStop}%
\bibitem [{\citenamefont {{Ali-Ha{\"\i}moud}}\ and\ \citenamefont
  {{Hirata}}(2010)}]{2010PhRvD..82f3521A}%
  \BibitemOpen
  \bibfield  {author} {\bibinfo {author} {\bibfnamefont {Y.}~\bibnamefont
  {{Ali-Ha{\"\i}moud}}}\ and\ \bibinfo {author} {\bibfnamefont {C.~M.}\
  \bibnamefont {{Hirata}}},\ }\bibfield  {title} {\bibinfo {title} {{Ultrafast
  effective multilevel atom method for primordial hydrogen recombination}},\
  }\href {https://doi.org/10.1103/PhysRevD.82.063521} {\bibfield  {journal}
  {\bibinfo  {journal} {\prd}\ }\textbf {\bibinfo {volume} {82}},\ \bibinfo
  {eid} {063521} (\bibinfo {year} {2010})},\ \Eprint
  {https://arxiv.org/abs/1006.1355} {arXiv:1006.1355 [astro-ph.CO]}
  \BibitemShut {NoStop}%
\bibitem [{\citenamefont {{Ali-Ha{\"\i}moud}}\ and\ \citenamefont
  {{Hirata}}(2011)}]{2011PhRvD..83d3513A}%
  \BibitemOpen
  \bibfield  {author} {\bibinfo {author} {\bibfnamefont {Y.}~\bibnamefont
  {{Ali-Ha{\"\i}moud}}}\ and\ \bibinfo {author} {\bibfnamefont {C.~M.}\
  \bibnamefont {{Hirata}}},\ }\bibfield  {title} {\bibinfo {title} {{HyRec: A
  fast and highly accurate primordial hydrogen and helium recombination
  code}},\ }\href {https://doi.org/10.1103/PhysRevD.83.043513} {\bibfield
  {journal} {\bibinfo  {journal} {\prd}\ }\textbf {\bibinfo {volume} {83}},\
  \bibinfo {eid} {043513} (\bibinfo {year} {2011})},\ \Eprint
  {https://arxiv.org/abs/1011.3758} {arXiv:1011.3758 [astro-ph.CO]}
  \BibitemShut {NoStop}%
\bibitem [{\citenamefont {{Fixsen}}\ \emph {et~al.}(2011)\citenamefont
  {{Fixsen}}, \citenamefont {{Kogut}}, \citenamefont {{Levin}}, \citenamefont
  {{Limon}}, \citenamefont {{Lubin}}, \citenamefont {{Mirel}}, \citenamefont
  {{Seiffert}}, \citenamefont {{Singal}}, \citenamefont {{Wollack}},
  \citenamefont {{Villela}},\ and\ \citenamefont
  {{Wuensche}}}]{2011ApJ...734....5F}%
  \BibitemOpen
  \bibfield  {author} {\bibinfo {author} {\bibfnamefont {D.~J.}\ \bibnamefont
  {{Fixsen}}}, \bibinfo {author} {\bibfnamefont {A.}~\bibnamefont {{Kogut}}},
  \bibinfo {author} {\bibfnamefont {S.}~\bibnamefont {{Levin}}}, \bibinfo
  {author} {\bibfnamefont {M.}~\bibnamefont {{Limon}}}, \bibinfo {author}
  {\bibfnamefont {P.}~\bibnamefont {{Lubin}}}, \bibinfo {author} {\bibfnamefont
  {P.}~\bibnamefont {{Mirel}}}, \bibinfo {author} {\bibfnamefont
  {M.}~\bibnamefont {{Seiffert}}}, \bibinfo {author} {\bibfnamefont
  {J.}~\bibnamefont {{Singal}}}, \bibinfo {author} {\bibfnamefont
  {E.}~\bibnamefont {{Wollack}}}, \bibinfo {author} {\bibfnamefont
  {T.}~\bibnamefont {{Villela}}},\ and\ \bibinfo {author} {\bibfnamefont
  {C.~A.}\ \bibnamefont {{Wuensche}}},\ }\bibfield  {title} {\bibinfo {title}
  {{ARCADE 2 Measurement of the Absolute Sky Brightness at 3-90 GHz}},\ }\href
  {https://doi.org/10.1088/0004-637X/734/1/5} {\bibfield  {journal} {\bibinfo
  {journal} {\apj}\ }\textbf {\bibinfo {volume} {734}},\ \bibinfo {eid} {5}
  (\bibinfo {year} {2011})},\ \Eprint {https://arxiv.org/abs/0901.0555}
  {arXiv:0901.0555 [astro-ph.CO]} \BibitemShut {NoStop}%
\bibitem [{\citenamefont {{Feng}}\ and\ \citenamefont
  {{Holder}}(2018)}]{2018ApJ...858L..17F}%
  \BibitemOpen
  \bibfield  {author} {\bibinfo {author} {\bibfnamefont {C.}~\bibnamefont
  {{Feng}}}\ and\ \bibinfo {author} {\bibfnamefont {G.}~\bibnamefont
  {{Holder}}},\ }\bibfield  {title} {\bibinfo {title} {{Enhanced Global Signal
  of Neutral Hydrogen Due to Excess Radiation at Cosmic Dawn}},\ }\href
  {https://doi.org/10.3847/2041-8213/aac0fe} {\bibfield  {journal} {\bibinfo
  {journal} {\apjl}\ }\textbf {\bibinfo {volume} {858}},\ \bibinfo {eid} {L17}
  (\bibinfo {year} {2018})},\ \Eprint {https://arxiv.org/abs/1802.07432}
  {arXiv:1802.07432 [astro-ph.CO]} \BibitemShut {NoStop}%
\bibitem [{\citenamefont {{Bowman}}\ \emph {et~al.}(2018)\citenamefont
  {{Bowman}}, \citenamefont {{Rogers}}, \citenamefont {{Monsalve}},
  \citenamefont {{Mozdzen}},\ and\ \citenamefont
  {{Mahesh}}}]{2018Natur.555...67B}%
  \BibitemOpen
  \bibfield  {author} {\bibinfo {author} {\bibfnamefont {J.~D.}\ \bibnamefont
  {{Bowman}}}, \bibinfo {author} {\bibfnamefont {A.~E.~E.}\ \bibnamefont
  {{Rogers}}}, \bibinfo {author} {\bibfnamefont {R.~A.}\ \bibnamefont
  {{Monsalve}}}, \bibinfo {author} {\bibfnamefont {T.~J.}\ \bibnamefont
  {{Mozdzen}}},\ and\ \bibinfo {author} {\bibfnamefont {N.}~\bibnamefont
  {{Mahesh}}},\ }\bibfield  {title} {\bibinfo {title} {{An absorption profile
  centred at 78 megahertz in the sky-averaged spectrum}},\ }\href
  {https://doi.org/10.1038/nature25792} {\bibfield  {journal} {\bibinfo
  {journal} {\nat}\ }\textbf {\bibinfo {volume} {555}},\ \bibinfo {pages} {67}
  (\bibinfo {year} {2018})},\ \Eprint {https://arxiv.org/abs/1810.05912}
  {arXiv:1810.05912 [astro-ph.CO]} \BibitemShut {NoStop}%
\end{thebibliography}%
\end{document}